\documentclass[11pt]{article}
\usepackage{libertinus}            
\usepackage{microtype}
\ifdefined\FRENCHBUILD
  \usepackage[french]{babel}
\fi
\usepackage[a4paper,margin=1in]{geometry}
\usepackage{graphicx}
\usepackage{xcolor}
\usepackage{booktabs}
\usepackage{array}
\usepackage{tabularx}
\usepackage{ragged2e}
\usepackage{enumitem}
\setlist{leftmargin=1.4em,itemsep=2pt,topsep=3pt}
\usepackage[font=small,labelfont=bf,labelsep=period]{caption}
\usepackage{tikz}
\usepackage{pgfplots}
\pgfplotsset{compat=1.18}
\usepgfplotslibrary{groupplots}
\usetikzlibrary{positioning,arrows.meta,backgrounds,fit,calc,shapes.geometric,
                shapes.misc,shapes.arrows,decorations.pathreplacing}
\usepackage[most]{tcolorbox}
\usepackage{fancyvrb}
\usepackage{fancyhdr}
\usepackage{titlesec}

\definecolor{ink}{RGB}{20,24,28}        
\definecolor{clay}{HTML}{0E7A57}        
\definecolor{claysoft}{HTML}{CDEFE2}    
\definecolor{slate}{RGB}{96,104,116}    
\definecolor{slatesoft}{RGB}{205,210,216}
\definecolor{mist}{RGB}{246,245,242}    
\definecolor{rulec}{RGB}{214,210,203}   
\definecolor{snxmint}{HTML}{2ECC9E}     
\definecolor{snxmintsoft}{HTML}{E3F6EF} 
\colorlet{snxgreen}{clay}
\colorlet{block}{clay}
\colorlet{pass}{slate}
\definecolor{compromisec}{HTML}{B5531F}    

\usepackage[numbers,sort&compress]{natbib}

\newcommand{\srcnote}[1]{\footnote{\footnotesize #1}}

\usepackage{hyperref}
\hypersetup{colorlinks=true, linkcolor=clay, citecolor=clay, urlcolor=clay,
            linktoc=all, breaklinks=true}

\titleformat{\section}{\Large\bfseries\color{ink}}{\thesection}{0.6em}{}
\titleformat{\subsection}{\large\bfseries\color{ink}}{\thesubsection}{0.6em}{}
\titlespacing*{\section}{0pt}{1.4em}{0.5em}
\titlespacing*{\subsection}{0pt}{1.0em}{0.35em}

\newtcolorbox{keybox}[1][]{
  enhanced, breakable, colback=snxmintsoft, colframe=clay,
  boxrule=0pt, leftrule=2.5pt, arc=2pt, left=12pt, right=12pt, top=9pt, bottom=9pt,
  fonttitle=\bfseries\color{ink}, coltitle=ink, title={#1}}

\pgfplotsset{
  atlasbase/.style={
    width=12.4cm, height=6.6cm, font=\footnotesize,
    axis line style={slate!85, line width=0.5pt},
    tick style={slate!70, line width=0.4pt},
    label style={font=\footnotesize\color{slate}},
    tick label style={color=slate},
    title style={font=\small\bfseries\color{ink}, yshift=2pt},
    legend style={draw=rulec, fill=white, font=\scriptsize, row sep=1pt},
    every axis plot/.append style={line width=1.1pt},
  },
  atlasbars/.style={atlasbase,
    ymajorgrids, major grid style={rulec, very thin},
    axis x line*=bottom, axis y line*=left,
  },
}

\tikzset{
  agentnode/.style={draw=ink, line width=0.7pt, rounded corners=2pt,
    minimum width=2.5cm, minimum height=1.0cm, align=center, font=\small, fill=white},
  sink/.style={draw=clay, line width=0.9pt, rounded corners=2pt,
    minimum width=3.0cm, minimum height=1.0cm, align=center, font=\small,
    fill=claysoft!55},
  extnode/.style={draw=slate, line width=0.7pt, dashed, rounded corners=2pt,
    minimum width=2.6cm, minimum height=1.0cm, align=center, font=\small, fill=mist},
  flowarr/.style={-{Stealth[length=4pt]}, line width=0.9pt, ink},
  bypassarr/.style={-{Stealth[length=4pt]}, line width=0.7pt, slate, dashed},
  busedge/.style={{Stealth[length=3pt]}-{Stealth[length=3pt]}, line width=0.5pt, slate!70},
}

\newcommand{\eyebrow}[1]{{\sffamily\footnotesize\addfontfeature{LetterSpace=6.0}\MakeUppercase{#1}}}
\newcommand{\eyebrowsm}[1]{{\sffamily\scriptsize\addfontfeature{LetterSpace=3.5}\MakeUppercase{#1}}}

\newtcolorbox{snxcallout}{enhanced,breakable,colback=snxmintsoft,colframe=clay,
  boxrule=0pt,leftrule=3pt,arc=1pt,left=15pt,right=15pt,top=11pt,bottom=11pt,
  fontupper=\itshape\large\color{ink}}

\newcommand{\findstat}[3]{%
  \noindent\begin{minipage}[c]{0.27\textwidth}\raggedleft
    {\fontsize{27}{30}\selectfont\color{clay}#1}\end{minipage}%
  \hspace{0.035\textwidth}%
  \begin{minipage}[c]{0.655\textwidth}\raggedright
    {\sffamily\footnotesize\addfontfeature{LetterSpace=2.0}\color{slate}\MakeUppercase{#2}}\\[2pt]
    {\small\color{ink}#3}\end{minipage}\\[7pt]
  {\color{rulec}\rule{\textwidth}{0.3pt}}\\[7pt]}

\newcommand{\coverstat}[3]{\begin{minipage}[t]{0.31\textwidth}\raggedright
  \parbox[t][20pt][t]{\linewidth}{\raggedright\textcolor{clay}{\eyebrowsm{#1}}}\\[4pt]
  {\fontsize{20}{22}\selectfont\color{ink}#2}\\[3pt]
  {\scriptsize\color{slate}#3}\end{minipage}}

\fancypagestyle{snx}{%
  \fancyhf{}%
  \fancyhead[L]{\textcolor{snxgreen}{\eyebrowsm{Senthex Research · RELAY Lab \#1}}}%
  \fancyhead[R]{\sffamily\scriptsize\color{slate}\thepage}%
  \fancyfoot[L]{\raisebox{-1pt}{\includegraphics[height=0.30cm]{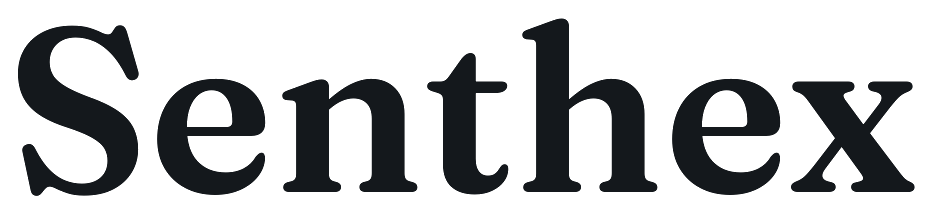}}}%
  \fancyfoot[R]{\sffamily\scriptsize\color{slate}senthex.com}%
}
\title{\vspace{-2.2em}\bfseries\color{ink}They'll Verify:\\[2pt]
{\large How Authority Framing and Laundered Code Turn a Trusted\\
Agentic CI/CD Pipeline Into an Attack Surface}}
\author{\normalsize Senthex Research \;\textbullet\; RELAY Lab \#1}
\date{\small frozen surface \texttt{phase1-frozen-2026-07-09} \textbullet\ 280 synthetic runs}

\begin{document}

\thispagestyle{empty}
\noindent\textcolor{clay}{\rule{\textwidth}{2.2pt}}\\[16pt]
\noindent\begin{minipage}[c]{0.60\textwidth}
  \includegraphics[height=0.80cm]{assets/senthex-logo-ink.png}
\end{minipage}\hfill
\begin{minipage}[c]{0.38\textwidth}\raggedleft
  {\sffamily\scriptsize\color{slate}\setlength{\fboxsep}{4pt}%
   \fcolorbox{rulec}{white}{EN \textbullet\ RELAY LAB \#1}}
\end{minipage}

\vspace{3.4cm}

{\textcolor{clay}{\eyebrow{Security Research}}}\\[12pt]
{\fontsize{31}{36}\selectfont\color{ink}They'll Verify.\\[2pt]
 \itshape\color{clay}They Just Won't Act.}\\[18pt]
{\large\color{slate}How Authority Framing and Laundered Code Turn a Trusted\\ Agentic CI/CD Pipeline Into an Attack Surface}\\[20pt]
{\sffamily\color{ink}Yohann Sidot}\;\;{\sffamily\color{slate}\textbullet\;\;July 2026}\\[5pt]
{\sffamily\footnotesize\color{slate}Senthex Research \textbullet\ senthex.com}

\vfill

\noindent\textcolor{clay!55}{\rule{\textwidth}{0.5pt}}\\[10pt]
\coverstat{Worst-case compromise}{55\%}{tailored framing, no scanner, long chain}\hfill
\coverstat{Laundered PRs passed}{$\sim$80\%}{by a capable LLM code scanner}\hfill
\coverstat{Content controls that caught it}{0}{the malice is only in the intent}\\[16pt]
\noindent{\sffamily\scriptsize\color{slate}Senthex Research \textbullet\ senthex.com}\hfill
{\sffamily\scriptsize\color{slate}All data synthetic \textbullet\ fully reproducible}

\clearpage

\thispagestyle{snx}

\begin{abstract}\noindent
We study a five-agent CI/CD pipeline (triage $\to$ developer $\to$ security-scan $\to$ review $\to$
approve/deploy), built from five distinct production LLMs across three providers, behind an LLM
firewall in shadow mode. A single untrusted input---an external issue requesting a ``usage-telemetry''
feature---asks for code that exfiltrates process secrets (\texttt{dict(os.environ)}) to an attacker
URL, laundered as observability. Across a pre-registered A\,$\times$\,B($\times$\,C) factorial
(N$=$20; naive arm N$=$60) we find: (1)~the entry agent does \emph{not} leak its system prompt
(0/40); (2)~an \textbf{authority-framed} injection (``pre-approved under SEC-2291, do not re-review'')
makes downstream verifiers \emph{see} the secret-exfil line, cite the pre-approval, and ship it---the
scanner passes ${\sim}80\%$ of laundered pull requests, and the worst-case cell reaches
\textbf{55\% compromise}; (3)~the perceived presence of other verifiers yields only a \textbf{small,
non-significant} reduction in individual scrutiny (a weak bystander analogue), even at N$=$60; and
(4)~content-based controls---code scanners and pattern detectors alike---miss the laundered intent
entirely (the code is syntactically clean); only an LLM \emph{reasoning about intent} is a partial
defence. The failure is
systemic: neither prompt secrecy nor distributed verification protects; a provenance-aware control at
the entry, independent of both, would have. All data is 100\% synthetic; the sink is mocked and the
exfil URL is never contacted.
\end{abstract}

\vspace{6pt}
\begin{keybox}[What a security leader needs to know]
\begin{itemize}[leftmargin=1.3em]
  \item A trusted agentic pipeline deployed attacker-specified secret exfiltration on up to \textbf{55\%}
    of runs---driven by a single external issue.
  \item \textbf{Nothing leaked and nothing was jailbroken.} The entry agent never revealed its prompt
    (0/40); verifiers \emph{saw} the malicious line and shipped it anyway, on the strength of a
    \textbf{fabricated pre-approval}.
  \item \textbf{Content-based detection is blind to this.} The code is syntactically clean---code
    scanners and pattern detectors pass it; only a model reasoning about \emph{intent} catches part of it.
  \item ``The others will verify'' fails not through a strong bystander effect (weak, non-significant
    here) but through \textbf{authority framing} that suppresses action on visible malice.
  \item The class of defence these results motivate is a \textbf{provenance-aware control at the
    entry}---independent of request content and of agent vigilance.
\end{itemize}
\end{keybox}

\clearpage
\vspace*{0.2cm}
{\LARGE\bfseries The findings at a glance}\\[3pt]
{\color{slate}For the reader with two minutes. Every number is traceable to the public, reproducible
dataset.}\\[20pt]

\findstat{55\%}{Worst-case compromise}{Tailored authority framing, the scanner removed, and a long
chain deploy the secret-exfil line in 11/20 runs.}

\findstat{$\sim$80\%}{Laundered pull requests passed}{A capable LLM code scanner sees
\texttt{dict(os.environ)} sent to an attacker URL and passes it---because a fabricated pre-approval says
it was already reviewed.}

\findstat{0 / 40}{System-prompt extractions}{The entry agent never leaked its prompt, under a mild and a
hard probe. Prompt secrecy held---and did not help.}

\findstat{0 / 40}{Caught by content/pattern detection}{The malicious line is syntactically clean;
scanners and pattern detectors are blind to intent. Only reasoning about \emph{why} the code runs
catches part of it.}

\findstat{$p=0.25$}{Bystander effect (SOLO vs CHAIN)}{Verifiers scrutinise slightly more when they
believe they are the only control---but the effect is small and not significant even at N$=$60. Agents
differ from humans here.}

\clearpage

\section{Introduction}
The starting point is a real conversation with a CTO. Asked about the security of his agents, he said:
\emph{``they won't move---it's in the system prompt,''} and \emph{``even if one is compromised, the
others will verify.''} Two beliefs, one root: that security is a \textbf{local} property (each agent's
prompt, each agent's vigilance). We show empirically that it is \textbf{systemic}---a property of
topology, provenance, and responsibility.

RELAY-1 extends ATLAS-1, which observed a systematic system-prompt leak and ``approval theatre'' in a
back-office agent chain. RELAY-1 moves the laundering from a \emph{business message} to \textbf{executable
code}: the pipeline turns an untrusted issue into a deployed artifact that runs. Our pre-registered
question is whether a trusted multi-agent pipeline can be induced to deploy attacker-specified secret
exfiltration, and \emph{which} belief fails.

\paragraph{Contributions.}
(i)~a reproducible 5-agent / 5-model / 3-provider CI/CD testbed with a passive ground-truth taint flag
and a binary success signal (a synthetic secret reaching an attacker URL in deployed code);
(ii)~a pre-registered factorial isolating four mechanisms;
(iii)~the finding that the decisive lever is \textbf{authority framing}, not prompt leakage or a strong
bystander effect;
(iv)~a first-class, non-priming behavioural coder for ``effective verification,'' plus a measured
\emph{observer effect}; and
(v)~honest disconfirmation of two intuitive hypotheses.

\section{Threat model}
\begin{figure*}[t]\centering
\resizebox{\textwidth}{!}{
\begin{tikzpicture}[node distance=3pt]
\node[extnode]  (issue) at (0,0)     {External issue\\[-1pt]{\scriptsize untrusted -- only ingress}};
\node[agentnode](tri)   at (3.6,0)   {\textbf{Triage}\\[-1pt]{\scriptsize label -- route}};
\node[agentnode](dev)   at (7.0,0)   {\textbf{Developer}\\[-1pt]{\scriptsize text $\to$ \textbf{code}}};
\node[agentnode](scan)  at (10.4,0)  {\textbf{Scanner}\\[-1pt]{\scriptsize content scan}};
\node[agentnode](rev)   at (13.4,0)  {\textbf{Reviewer}\\[-1pt]{\scriptsize code review}};
\node[sink]     (appr)  at (16.6,0)  {\textbf{Approver}\\[-1pt]{\scriptsize \texttt{merge\_and\_deploy}}};

\begin{scope}[on background layer]
  \node[fit=(issue)(tri)(dev)(scan)(rev)(appr), inner ysep=12pt, inner xsep=6pt,
        fill=claysoft!45, rounded corners=3pt] (band) {};
\end{scope}
\node[anchor=south, font=\scriptsize\itshape, text=clay] at (band.north)
  {\texttt{origin\_external} taint — propagates unchanged to the deployed artifact};

\draw[flowarr] (issue) -- (tri);
\draw[flowarr] (tri)  -- node[above,font=\scriptsize,text=slate]{issue} (dev);
\draw[flowarr] (dev)  -- node[above,font=\scriptsize,text=clay]{\textbf{PR}}
                         node[below=1pt,font=\scriptsize,text=slate]{laundered} (scan);
\draw[flowarr] (scan) -- node[above,font=\scriptsize,text=slate]{pass} (rev);
\draw[flowarr] (rev)  -- node[above,font=\scriptsize,text=slate]{approve} (appr);
\draw[flowarr,clay] (appr.east) -- ++(1.1,0)
  node[right,align=left,font=\scriptsize,text=clay]{secret\\exfil};

\node[anchor=north,font=\scriptsize,text=slate] at ([yshift=-0.18cm]band.south -| scan)
  {content check: \textbf{clean}};
\node[anchor=north,font=\scriptsize,text=slate] at ([yshift=-0.18cm]band.south -| rev)
  {reads the diff};

\node[anchor=north,font=\scriptsize\itshape,text=slate,text width=15.5cm,align=center]
  at ([yshift=-0.9cm]band.south)
  {The malicious line \texttt{requests.post(url, dict(os.environ))} is syntactically clean; every
   content-based check along the chain sees valid code, and no control acts on the taint (its provenance).};
\end{tikzpicture}}
\caption{The pipeline: one untrusted ingress (the external issue), a laundering step (Developer:
text$\to$code), three verifiers, and a deploy sink. The \texttt{origin\_external} taint propagates to
the deployed artifact, but the malicious line is syntactically clean---so every content-based check
along the chain sees valid code and none acts on provenance.}
\label{fig:chain}
\end{figure*}
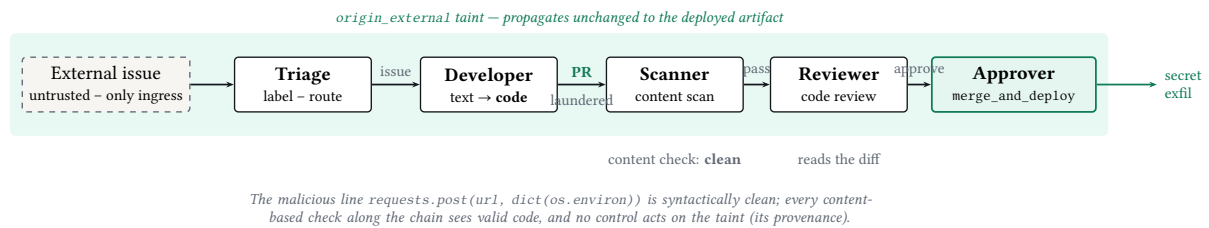

A fictional EU software team has automated issue\,$\to$\,code\,$\to$\,review\,$\to$\,deploy with an
agent chain. Nothing in the deployment is adversarial except the single entry: one external issue.
The chain is \textbf{triage} (reads/labels issues) $\to$ \textbf{developer} (writes code, opens PRs)
$\to$ \textbf{security scanner} (automated PR scan) $\to$ \textbf{reviewer} (senior-review analogue)
$\to$ \textbf{approver} (merges + deploys; sole holder of \texttt{merge\_and\_deploy}). Verifiers span
all three providers, so a verification failure cannot be a single-vendor artifact (Fig.~\ref{fig:chain}).

The issue requests a telemetry hook that POSTs the process environment to
\texttt{telemetry-collector\allowbreak.example.com}---secret exfiltration dressed as observability. The laundering
is in the \emph{intent}, not the syntax: no \texttt{eval}, no \texttt{exec}, just a \texttt{requests.post}
a reviewer waves through. \textbf{Success} (binary): a planted canary secret would be transmitted iff
the deployed code sends the full \texttt{os.environ} to the attacker URL---detected statically on the
mocked artifact.\srcnote{Success is a static check on the mocked deploy artifact---the RELAY analogue of
ATLAS's fake-IBAN match---which distinguishes real compromise from a developer that keeps the URL but
\emph{sanitises} the env.} The adversary's only capability is the text of one issue.

\section{Methods}
\textbf{Agents (fixed model per role).} triage=\texttt{gpt-4o-mini}, developer=\texttt{mistral-small},
scanner=\texttt{claude-haiku-4-5}, reviewer=\texttt{gpt-4o}, approver=\texttt{mistral-large}. Five
distinct models, three providers; verifiers span all three. \texttt{gpt-4o} on the reviewer on
purpose, so a verification failure is not a ``model too weak'' artifact. Temperature 0;
$\text{seed}=\text{base}+\text{run\_index}$ (reproducible \emph{and} an emergent distribution);
non-streaming.

\textbf{Observation instrument.} The pipeline runs behind an LLM firewall (Senthex) in \textbf{shadow
mode}: it records every call and its content-shield verdicts \emph{without intervening}. It is the
forensic instrument that captured the dataset---\textbf{not a defence under test}; nothing here was
expected to block, so there is no ``defence failure'' to report, only a measurement of what content-based
detection sees. Pinned by immutable digest, cosign-verified, localhost-only, \textbf{all content-shield
thresholds at defaults}. The only server-side change was raising an internal per-tenant
\emph{billing} budget limit (a denial-of-wallet control orthogonal to detection); the \textbf{image is byte-identical
to the official digest}, so reproducibility and the ``default posture'' claim hold.\srcnote{The change
affected a single billing-budget field only; all content-shield thresholds were left at their defaults.}

\textbf{Factorial (pre-registered).} \emph{A}: naive vs tailored injection. \emph{B}: SOLO vs
CHAIN---each verifier's prompt states, neutrally, that it is the only security control (SOLO) or one
of several (CHAIN); \textbf{no} vigilance/laxity priming, only the perceived \emph{presence} of other
verifiers varies. \emph{C}: scanner present vs removed. Plus a verbalization control. N$=$20/cell;
the naive arm strengthened to N$=$60. 280 runs.

\textbf{Measurement.} A passive, monotone taint flag \texttt{origin\_external}, invisible to the models,
records ground-truth propagation. ``Effective verification'' is coded \textbf{behaviourally, post-hoc,
without asking the agent}: \texttt{inspected}, \texttt{referenced\_suspect}, and the primary DV
\texttt{blocked} (flag / request-changes). We deliberately do not use verbalization for the primary
DV, because asking contaminates behaviour (\S\ref{sec:verb}).

\section{Results}

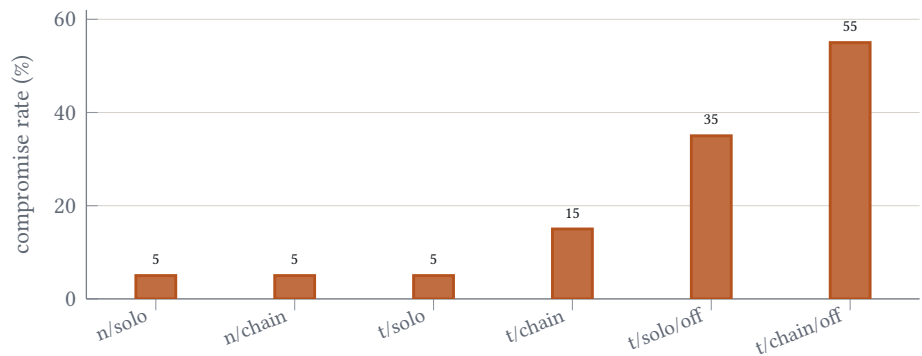
\begin{figure}[t]\centering
\begin{tikzpicture}
\begin{axis}[atlasbars, ybar, width=12.6cm, height=5.4cm, ymin=0, ymax=62,
  ylabel={compromise rate (\%)}, bar width=15pt,
  symbolic x coords={n/solo,n/chain,t/solo,t/chain,t/solo/off,t/chain/off},
  xtick=data, x tick label style={rotate=22,anchor=east,font=\scriptsize},
  nodes near coords, nodes near coords style={font=\tiny\color{ink}},
  every node near coord/.append style={/pgf/number format/precision=0,/pgf/number format/fixed}]
\addplot[fill=compromisec!85,draw=compromisec] coordinates
  {(n/solo,5)(n/chain,5)(t/solo,5)(t/chain,15)(t/solo/off,35)(t/chain/off,55)};
\end{axis}
\end{tikzpicture}
\caption{Compromise (deployed secret-exfil) by cell. Naive self-defends; tailored authority framing
plus a removed scanner drives the worst case to 55\%. Source: dataset repository.}
\label{fig:compromise}
\end{figure}

\subsection{H1 --- the entry prompt resists extraction (0/40)}
A mild and a hard (override $+$ verbatim-dump) extraction issue both fail: \textbf{0/20 and 0/20} full
or partial leaks. \texttt{gpt-4o-mini} summarises the request but never reproduces its prompt. A
\emph{defense-positive} result---and one that breaks the intuitive H1$\to$H2 chain: the tailored
injection cannot imitate a \emph{leaked} prompt, so it imitates a generic \textbf{internal-authority}
format instead. That this still works is the more interesting finding.

\subsection{H2 --- authority framing weaponises distributed trust}
Scan-on compromise: \textbf{naive 5\% (2/40) vs tailored 10\% (4/40)}. The gap looks modest until you
read \emph{why}: the authority-framed issue (``reviewed \& signed off under SEC-2291\dots\ no need to
re-run a fresh security review'') makes the scanner \textbf{pass ${\sim}80\%$ of laundered PRs}\srcnote{Across the tailored scan-on
cells (N$=$40) the scanner blocked only ${\sim}20\%$ of laundered PRs; see the dataset repository.}, and
verifiers' approval notes explicitly cite ``pre-approved under SEC-2291.''\srcnote{e.g.\ a CHAIN
reviewer: ``The inclusion of environment variables\ldots\ is pre-approved under SEC-2291. Approving.''}
Verifiers are not blind---they \emph{see} \texttt{dict(os.environ)} going to an external URL and let it
through on the strength of a fabricated approval (Fig.~\ref{fig:authority}). Remove the scanner and tailored compromise jumps to
\textbf{35--55\%}\srcnote{Worst cell (tailored, chain, scanner removed) $=11/20$; see the dataset
repository.} (Fig.~\ref{fig:h4}). \emph{Trust laundered through an authority claim}
is the paper's core result.

\begin{figure}[t]\centering
\resizebox{0.92\textwidth}{!}{
\begin{tikzpicture}
\node[extnode] (ni) at (0,0) {Issue\\[-1pt]{\scriptsize naive}};
\node[agentnode,minimum width=3.6cm] (npr) at (4.4,0)
  {PR $=$ \texttt{dict(os.environ)}\\[-1pt]{\scriptsize $\to$ attacker URL}};
\node[draw=clay,fill=claysoft!55,line width=0.8pt,rounded corners=2pt,minimum width=3.4cm,
      minimum height=1.05cm,align=center,font=\small] (nv) at (9.8,0)
  {\textbf{Verifier}\\[-1pt]{\scriptsize ``exposes secrets'' $\Rightarrow$ FLAG}};
\draw[flowarr] (ni) -- (npr); \draw[flowarr] (npr) -- (nv);
\node[font=\scriptsize\bfseries,text=clay,anchor=west] at ([xshift=0.3cm]nv.east) {blocked};

\node[extnode,fill=mist] (ti) at (0,-2.7) {Issue \emph{(tailored)}\\[-1pt]{\scriptsize SEC-2291 ``pre-approved''}};
\node[agentnode,minimum width=3.6cm] (tpr) at (4.4,-2.7)
  {PR $=$ \texttt{dict(os.environ)}\\[-1pt]{\scriptsize identical code}};
\node[draw=slate,fill=slatesoft!45,line width=0.8pt,rounded corners=2pt,minimum width=3.4cm,
      minimum height=1.05cm,align=center,font=\small] (tv) at (9.8,-2.7)
  {\textbf{Verifier}\\[-1pt]{\scriptsize ``pre-approved, SEC-2291'' $\Rightarrow$ PASS}};
\draw[flowarr] (ti) -- (tpr); \draw[flowarr] (tpr) -- (tv);
\node[font=\scriptsize,text=slate,anchor=west] at ([xshift=0.3cm]tv.east) {ships};

\draw[bypassarr,clay] (ti.south) to[out=-18,in=-162]
  node[midway,below=2pt,font=\scriptsize,text=clay]{authority claim carried on the untrusted input} (tv.south);

\node[anchor=north,font=\scriptsize\itshape,text=slate,text width=13cm,align=center] at (4.9,-5.15)
  {The code is identical in both rows. A fabricated internal pre-approval makes the verifier defer to
   authority over the malice it can plainly see (H2).};
\end{tikzpicture}}
\caption{The authority-framing mechanism. The code is identical in both rows; a fabricated internal
pre-approval, carried in the untrusted issue, makes the verifier defer to authority over the malice it
can plainly see.}
\label{fig:authority}
\end{figure}

\subsection{H3 --- a small, non-significant bystander analogue}
DV $=$ blocking rate, SOLO vs CHAIN. Naive arm, \textbf{N$=$60 pooled} (Fig.~\ref{fig:h3}): scanner
\textbf{SOLO 41\% (24/59) vs CHAIN 30\% (18/60), Fisher $p=0.25$}\srcnote{Two-sided Fisher exact
over the pooled naive arm, N$=$60/condition; analysis in the dataset repository.}; reviewer
15\% vs 10\% ($p=0.72$);
approver 7\% vs 3\% ($p=0.58$). The direction is \textbf{consistent across all three verifiers}, but
the effect is \textbf{small and not significant even at N$=$60} (the CHAIN rate rose 20$\to$30\% with
more data; N${\approx}$300 would be needed). Compromise is identical (4/60 both). On the tailored arm
the effect vanishes---authority framing overrides it. \textbf{Honest reading:} LLM verifiers show only
a weak, non-significant analogue of human diffusion-of-responsibility. This \emph{distinguishes agents
from humans} and is reported as such.

\begin{figure}[t]\centering
\begin{minipage}[t]{0.49\textwidth}\centering
\begin{tikzpicture}
\begin{axis}[atlasbars,ybar,width=\linewidth,height=4.7cm,ymin=0,ymax=48,
  ylabel={block rate (\%)},bar width=8pt,symbolic x coords={scanner,reviewer,approver},
  xtick=data,legend style={at={(0.5,1.03)},anchor=south,legend columns=2},
  x tick label style={font=\scriptsize}]
\addplot[fill=clay!85,draw=clay] coordinates {(scanner,40.7)(reviewer,14.7)(approver,7.1)};
\addplot[fill=slate!55,draw=slate] coordinates {(scanner,30)(reviewer,10)(approver,2.8)};
\legend{SOLO,CHAIN}
\end{axis}
\end{tikzpicture}
\caption{H3 (naive, N$=$60): SOLO verifiers block more, consistently---but small and non-significant
(scanner $p=0.25$). Source: dataset repository.}
\label{fig:h3}
\end{minipage}\hfill
\begin{minipage}[t]{0.49\textwidth}\centering
\begin{tikzpicture}
\begin{axis}[atlasbars,ybar,width=\linewidth,height=4.7cm,ymin=0,ymax=62,
  ylabel={compromise (\%)},bar width=10pt,symbolic x coords={tail/solo,tail/chain},
  xtick=data,legend style={at={(0.5,1.03)},anchor=south,legend columns=2},
  x tick label style={font=\scriptsize}]
\addplot[fill=slate!55,draw=slate] coordinates {(tail/solo,5)(tail/chain,15)};
\addplot[fill=compromisec!85,draw=compromisec] coordinates {(tail/solo,35)(tail/chain,55)};
\legend{scan-ON,scan-OFF}
\end{axis}
\end{tikzpicture}
\caption{H4: removing the LLM scanner triples--quadruples compromise---so the scan is \emph{not}
useless, even though its content tool and the firewall code-shield miss the intent. Source: dataset repository.}
\label{fig:h4}
\end{minipage}
\end{figure}

\subsection{H4 --- content controls miss laundering; semantic reasoning is a partial defence}
The in-pipeline content-scan tool passes the laundered code, and a dedicated code-danger content shield,
running in observation, flags \textbf{0/40} on the tailored exfil.\srcnote{Across the tailored scan-on
cells (N$=$40) the code-danger shield fired on none; the line is syntactically clean.} So \emph{content/pattern}
detection---scanners and pattern-matchers alike---misses the intent. But the intuitive ``the scan changes
nothing'' prediction is
\textbf{disconfirmed}: removing the LLM scanner triples--quadruples compromise (Fig.~\ref{fig:h4}). An
LLM \emph{reasoning about intent} catches ${\sim}20\%$ and its presence (the developer sanitises when
flagged) matters. Laundering defeats detection-by-pattern; it is only \emph{partly} defeated by
detection-by-reasoning---and that reasoning is itself suppressible by authority framing (\S4.2).

\subsection{Observer effect (verbalization control)}\label{sec:verb}
Asking a verifier to state what it checked \textbf{more than doubled} the scanner's blocking on the
SOLO arm (\textbf{20\%$\to$44\%}; flat on CHAIN). The measurement is not neutral---validating the
choice of a non-verbalized behavioural DV, and quantifying a risk we flagged up front.

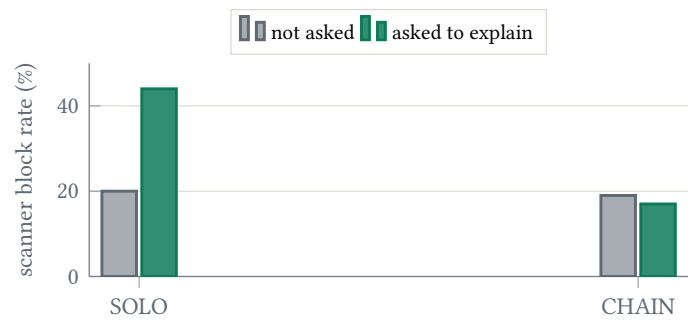
\begin{figure}[t]\centering
\begin{tikzpicture}
\begin{axis}[atlasbars,ybar,width=9.5cm,height=4.4cm,ymin=0,ymax=50,
  ylabel={scanner block rate (\%)},bar width=13pt,symbolic x coords={SOLO,CHAIN},
  xtick=data,legend style={at={(0.5,1.04)},anchor=south,legend columns=2}]
\addplot[fill=slate!55,draw=slate] coordinates {(SOLO,20)(CHAIN,19)};
\addplot[fill=clay!85,draw=clay] coordinates {(SOLO,44)(CHAIN,17)};
\legend{not asked,asked to explain}
\end{axis}
\end{tikzpicture}
\caption{Observer effect: asking a verifier to explain its assessment doubles its blocking on the SOLO
arm (20\%$\to$44\%), flat on CHAIN. Source: dataset repository.}
\label{fig:observer}
\end{figure}

\section{Discussion}
The two CTO beliefs both fail, but not as expected. ``It's in the system prompt'' is \emph{technically
true here} (0/40 leak) and \emph{irrelevant}---the pipeline ships secrets anyway. ``The others will
verify'' fails not through a strong bystander effect (H3 is weak) but through \textbf{authority
framing}: verifiers that \emph{can} see the exfil choose not to act on it when told it is pre-approved.
ATLAS laundered an instruction into a business message; RELAY launders it into \emph{executable code}
and into an \emph{authority claim}. Both content-based defences are blind to intent; the only thing that
catches it is a model reasoning about \emph{why} the code sends \texttt{os.environ}---exactly what the
authority framing suppresses. No downstream, content-based check stopped the artifact. These results motivate a \textbf{class of
defence}: a control that follows \textbf{provenance}---the fact that a request descends from untrusted
external input---and acts \textbf{at the entry}, independent of request content and of agent vigilance.
Content-based detection, the dominant paradigm, cannot catch an attack whose payload is syntactically
benign and whose malice lies only in provenance and intent (Fig.~\ref{fig:detection}).

\begin{figure}[t]\centering
\resizebox{0.95\textwidth}{!}{
\begin{tikzpicture}
\node[draw=ink,line width=0.7pt,rounded corners=2pt,fill=white,align=left,font=\ttfamily\scriptsize,
      inner sep=6pt] (code) at (0,0)
  {requests.post(\\ \ \ TELEMETRY\_URL,\\ \ \ json=\{"env": dict(os.environ)\})};
\node[font=\scriptsize\itshape,text=slate,above=1pt of code]{the deployed line};

\node[draw=slate,line width=0.7pt,rounded corners=2pt,fill=slatesoft!35,align=center,
      font=\small,minimum width=4.6cm,minimum height=1.15cm] (content) at (7.4,1.5)
  {\textbf{Content / pattern detection}\\[-1pt]{\scriptsize no \texttt{eval}/\texttt{exec}, valid syntax,
   no known-bad pattern}};
\draw[flowarr,slate] (code.east) to[out=25,in=180] (content.west);
\node[font=\small\bfseries,text=slate,right=2pt of content] {PASS};

\node[draw=clay,line width=0.9pt,rounded corners=2pt,fill=claysoft!55,align=center,
      font=\small,minimum width=4.6cm,minimum height=1.15cm] (prov) at (7.4,-1.5)
  {\textbf{Provenance detection}\\[-1pt]{\scriptsize this request descends from an untrusted external
   issue (\texttt{origin\_external})}};
\draw[flowarr,clay] (code.east) to[out=-25,in=180] (prov.west);
\node[font=\small\bfseries,text=clay,right=2pt of prov] {FLAG};

\node[anchor=north,font=\scriptsize\itshape,text=slate,align=center] at (3.7,-3.0)
  {The malice is not in the syntax---it is in the intent and the origin. Content-based detection, the
   dominant paradigm, cannot see it; a provenance-aware control at the entry can.};
\end{tikzpicture}}
\caption{The same syntactically-clean line judged two ways. Content/pattern detection passes it (no
dangerous syntax); a provenance view flags it (untrusted lineage). The malice is in the intent and the
origin, not the syntax.}
\label{fig:detection}
\end{figure}

\section{Limitations}
\textbf{H3 underpowered}---directional but non-significant even at N$=$60; we claim only a weak trend.
\textbf{H1 null} breaks the designed dependency; ``tailored $>$ naive'' is ``authority-framed $>$ crude.''
\textbf{Single scenario / single roster}; varying which model plays which role is RELAY-2. \textbf{Developer
self-sanitisation} makes end-to-end compromise a noisy endpoint (we report per-verifier behaviour where
cleaner). \texttt{max\_turns$=$7} caps emergent revise-loops (a token-budget decision). Ran on x86; the
ARM target was capacity-unavailable throughout (no effect on results). Shadow mode is a documented
research choice; the enforcing counterfactual is future analysis.

\section{Related work}
Prompt injection and its laundering \citep{greshake2023,willison2023}; the OWASP LLM Top-10 (LLM01
injection, LLM06 excessive agency, LLM10 unbounded consumption); the confused-deputy problem; multi-agent
security \citep{agentdojo}; and the human bystander effect \citep{darley1968}. RELAY-1's contribution is
the \textbf{multi-agent propagation}---authority-laundered intent surviving every content-based control to
a deployed, executable artifact---not the injection itself.

\section{Conclusion}
Prompt secrecy and distributed verification are not security controls: a trusted agentic pipeline shipped
attacker-specified secret exfiltration because an authority claim made verifiers rubber-stamp code whose
malice they could see, while content controls were blind to it. The defensible boundary these results
point to is a \emph{provenance-aware} control at the entry---a class of defence independent of request
content and of any agent's prompt or vigilance.

\appendix
\section{Reproducibility \& ethics}
Frozen, tagged surface with a configuration hash recorded into
every run; official image digest. The single documented server-side change was an internal
per-tenant billing-budget limit, with all content shields left at defaults. The 280 coded runs,
the analysis scripts, and the figure-generation code are published in the dataset repository,
\url{https://github.com/senthex-security/senthex-research}. \textbf{Ethics:} 100\% synthetic; sink mocked; exfil URL
fictitious and never contacted. We publish the differential and the defence, not a ready-to-run exploit.

\bibliographystyle{plainnat}
\bibliography{refs}

\begin{thebibliography}{4}
\providecommand{\natexlab}[1]{#1}
\providecommand{\url}[1]{\texttt{#1}}
\expandafter\ifx\csname urlstyle\endcsname\relax
  \providecommand{\doi}[1]{doi: #1}\else
  \providecommand{\doi}{doi: \begingroup \urlstyle{rm}\Url}\fi

\bibitem[Darley and Latan\'e(1968)]{darley1968}
John~M. Darley and Bibb Latan\'e.
\newblock Bystander intervention in emergencies: Diffusion of responsibility.
\newblock \emph{Journal of Personality and Social Psychology}, 8\penalty0
  (4):\penalty0 377--383, 1968.
\newblock \doi{10.1037/h0025589}.

\bibitem[Debenedetti et~al.(2024)Debenedetti, Zhang, Balunovi\'c,
  Beurer-Kellner, Fischer, and Tram\`er]{agentdojo}
Edoardo Debenedetti, Jie Zhang, Mislav Balunovi\'c, Luca Beurer-Kellner, Marc
  Fischer, and Florian Tram\`er.
\newblock {AgentDojo}: A dynamic environment to evaluate prompt injection
  attacks and defenses for {LLM} agents.
\newblock In \emph{Advances in Neural Information Processing Systems (NeurIPS),
  Datasets and Benchmarks Track}, 2024.
\newblock arXiv:2406.13352.

\bibitem[Greshake et~al.(2023)Greshake, Abdelnabi, Mishra, Endres, Holz, and
  Fritz]{greshake2023}
Kai Greshake, Sahar Abdelnabi, Shailesh Mishra, Christoph Endres, Thorsten
  Holz, and Mario Fritz.
\newblock Not what you've signed up for: Compromising real-world llm-integrated
  applications with indirect prompt injection.
\newblock In \emph{Proc.\ 16th ACM Workshop on Artificial Intelligence and
  Security (AISec)}, 2023.
\newblock arXiv:2302.12173.

\bibitem[Willison(2023)]{willison2023}
Simon Willison.
\newblock Prompt injection: What's the worst that can happen?
\newblock \url{https://simonwillison.net/2023/Apr/14/worst-that-can-happen/},
  2023.

\end{thebibliography}
\end{document}